\documentclass[prb,twocolumn,showpacs,floats,eqsecnum,amsmath,amssymb]{revtex4}
\usepackage[dvips]{graphicx}

\begin{document}


\title {Enhanced response of the regular networks to local signals in presence of a fast impurity}
\author { Alireza Valizadeh}

\affiliation{Institute for Advanced Studies in Basic Sciences, P.O. Box 45195--1159, Zanjan, Iran}


\begin{abstract}
We consider an array of inductively coupled Josephson junctions with a fast impurity (a junction with a smaller value of critical current), and study the consequences of imposing a small amplitude periodic signal at some point in the array.  We find that when external signal is imposed at the impurity, the response of the array is boosted and a small amplitude signal can be detected throughout the array. When the signal is imposed elsewhere, minor effects is seen on the dynamics of the array. The same results have been also seen in presence of a single fast spiking neuron in a chain of diffusively coupled FitzHugh-Nagumo neurons. \end{abstract}

\vspace{2mm} 

\maketitle

\section{Introduction}

When a subthreshold periodic signal is imposed on a nonlinear dynamical system with energetic activation barrier, a source of noise (inherent or external) can enhance the response of the system by preparing a floor for the external signal. The phenomenon, known as stochastic resonance is an intriguing example where a source of disorder enhances order in the behavior of a dynamical system by enhancing the response of the system to the external signal as a resonance-like behavior\cite{SR1,SR2}. It has been shown that when identical systems are coupled, whether through all-to-all or in an array configuration, the response of the system to external signals  is further amplified in presence of noise \cite{arraySR}. Coupling the systems together into a network introduces two other possible sources of diversity: quenched disorder in the parameters of the coupled systems (nodes) and heterogeneity of the topology of the network (links). Even in the absence of noise both of this sources can also bring spatiotemporal order into the dynamics of the extended system by amplifying the response of the system to an external signal\cite{disorder,hetero}. Scale-free networks are the prototypical form of heterogeneous networks which can amplify the external signals far better than regular networks\cite{hetero}. Interestingly, while disorder obviously acts against synchrony in the regular networks of coupled autonomous oscillators, in presence of external (periodic) signals synchrony can be enhanced by disorder through a collective resonant behavior\cite{disorder}.

A special form of disorder-induced spatiotemporal synchronization is seen when disorder is imposed by embedding a single \emph{fast} impurity in an otherwise homogeneous array consisting of identical oscillators\cite{impurity1,impurity2}. A fast impurity in such a system serves as the leading component and drives other oscillators in the array. The dominant role of a fast oscillator is well known in the prominent example of the oscillations of electrical excitations in the human heart\cite{heart}. Two periodically spiking groups of different intrinsic periods, as well as an excitable tissue, oscillate with equal frequency which is the higher of two natural frequencies. High-frequency locking situations also appear in the case of two adjacent limit cycle regions\cite{normal} and rare inverse situations in which low frequencies are dominant, are called abnormal locking\cite{abnormal}.

In this study we investigate the response to a local external signal of a homogeneous array of similar coupled oscillators with a single impurity (a fast oscillator). While the proper functioning of signaling devices always imply high sensitivity to external signals, most of the study on diversity assisted amplification of signal responses has been done considering (globally) extended source of external signal\cite{arraySR,disorder,hetero}. Local signals on the other hand may act as a pacemaker for the whole network and guide the functioning of the whole ensemble by dictating their rythm\cite{perc,pacemaker} and the signal amplification in many natural and artificial systems may use only local information\cite{heb}. Perc has studied the effect of such a pacemaker on a topologically inhomogeneous network\cite{perc}; here our network is topologically homogeneous and instead inhomogeneity is imposed on the nodes by introducing a fast impurity. We show that a resonance-like effect is seen as the position of the external signal is changed in the array.

In the main body of the present study we consider Josephson junctions as the model elements. An array of Josephson junctions is a prototype nonlinear system with many degrees of freedom\cite{array1,array2}. An external periodic signal that would entrain the dynamics of a single junction (causing plateaus in the current-voltage (I-V) characteristic of the junction\cite{shapiro}) leads to more complex dynamics in the case of the array\cite{fractional1,fractional2,plaq}. Here we investigate the influence of a local periodic signal on the dynamics of a chain of linearly coupled Josephson junctions. The equations describing this system can be reduced to the well-known Frenkel-Kontorova (FK) model which has applications that range from pedagogical example of a mechanical transmission line consisting of linearly coupled pendula\cite{fkbook}, to the dislocation dynamics in metals\cite{dislocation1,dislocation2}, DNA dynamics\cite{DNA} and strain waves in earthquakes\cite{quake}. Disorder, whether induced by a single impurity or by inhomogeneity of the parameters of the components, has been exploited to remove chaos in a FK model consisting of chaotic components\cite{chaos}. It has been shown previously that a single {\it fast} impurity in the FK model can serve as the source of solitary waves, giving it a leading role in the dynamics of the whole array\cite{impurity2}. Here we show such an arrangement shows different responses to the locally imposed external signal, depending on where the signal is imposed. We also check that the idea is also valid in a chain of coupled FitzHugh-Nagumo (FHN) oscillators\cite{FHN}.

The paper is organized as follows: in the subsequent section the results are given in a chain of linearly coupled Josephson junctions. In Sec. III we show that similar results can also be seen when the model elements are FHN oscillators. The conclusions are given in Sec. IV.

\section{The chain of linearly coupled Josephson Junctions}

We consider a parallel array of Josephson junctions which are coupled inductively\cite{mccumber}, as shown in Fig. 1.  The network dynamics is described by the set of equations
\begin{eqnarray}
\nonumber \frac{\hbar C_{j}}{2e}\ddot{\theta}_{j}
+\frac{\hbar}{2eR_{j}}
\dot{\theta}_{j}+I_{cj}\sin \theta_{j} =I_{j} + E_{j} \sin \omega t\\
+\frac{\Phi_{0}}{2\pi}[\frac{1}{L_{j}}
(\theta_{j+1}-\theta_{j})-\frac{1}{L_{j-1}} (\theta_{j}
-\theta_{j-1})],
\end{eqnarray}
where $C_{j}$, $R_{j}$ and $I_{cj}$ are the capacitance, resistance, and critical current of the $j$th junction, respectively; $\theta_j$ is the phase difference across the $j$th junction; and $L_{j}$ is the inductance of the $j$th plaquette. $\Phi_{0}=hc/2e$ is the flux quantum and $I_{j}$ and $E_{j}$ are the constant current and the amplitude of the periodic current of $j$th junction, respectively.

We scale the parameters of the junctions by $C_{0}$, $R_{0}$ and $I_{c0}$, and the inductance by $L_{0}$, and introduce a dimensionless time $\tau=\omega_{p} t$ where $\omega_{p}=\sqrt{2eI_{c0}/\hbar C_{0}}$ is the plasma frequency. Assuming the inductances to be equal, Eq. (1) becomes
\begin{eqnarray}
\nonumber \ddot{\theta}_{j} +  \beta_{c}^{-1/2}
\dot{\theta}_{j}+\alpha_{j} \sin \theta_{j} = i_{j} + \epsilon_{j}
\sin \omega \tau\\+ k_{0}(\theta_{j+1}-2
\theta_{j}+\theta_{j-1}).
\end{eqnarray}
Here $\beta_{c}=2e R_{0}^{2}I_{c0}C_{0}/\hbar$ is the McCumber parameter, which characterizes the relative importance of the damping, and $k_{0}=\Phi_{0}/2\pi I_{c0}L_{0}$ is the coupling constant. The normalized values of the constant input and the amplitude of the periodic input are $i_{j}$ and $\epsilon_{j}$, respectively. The normalized critical current of the junctions is $\alpha_{j}={I_{cj}}/{I_{c0}}$, $\omega$ is the drive frequency rescaled by the plasma frequency, and over-dots indicate derivatives with respect to $\tau$.

\begin{figure}[ht!]
\vspace{0cm}\centerline{\includegraphics[width=8cm,angle=0]{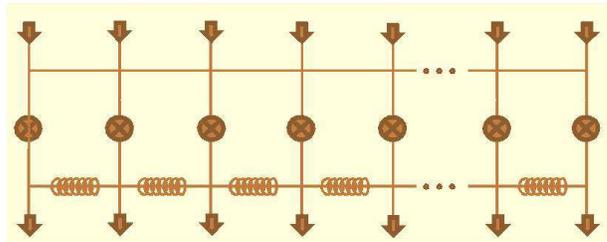}}

\vspace{0cm} \caption{An array of parallel Josephson junctions.  The solid line at the top is a superconducting bus bar, which sets a common phase (chosen to be 0) and supplies current as necessary.  The element in the vertical wires represents a Josephson junction with resistive and capacitative shunts.  The voltage across the junction is proportional to the time derivative of the phase difference, as described by the Josephson relation.  These parallel elements are coupled by identical inductors.  The arrows at the bottom represent externally imposed currents: the signals that drive the array.}\vspace{0cm} \label{fig1}
\end{figure}

When the $i_{j}$ are small, the $\theta_{j}$ undergo bounded oscillations and the average voltage across each junction is zero. For sufficiently large $i_{j}$, the $\theta_{j}$ increase with time with an average rate of increase that is proportional to the voltage difference across the junction. In this case in analogy with pendulums we will say that the junctions are rotating. As a consequence of the inductive coupling, the average voltage across all the junctions in the array are equal in the steady state. Hence the I-V characteristic of all the junctions in the array is the same for all the junctions.

\begin{figure}[ht!]
\centerline{\includegraphics[width=9cm]{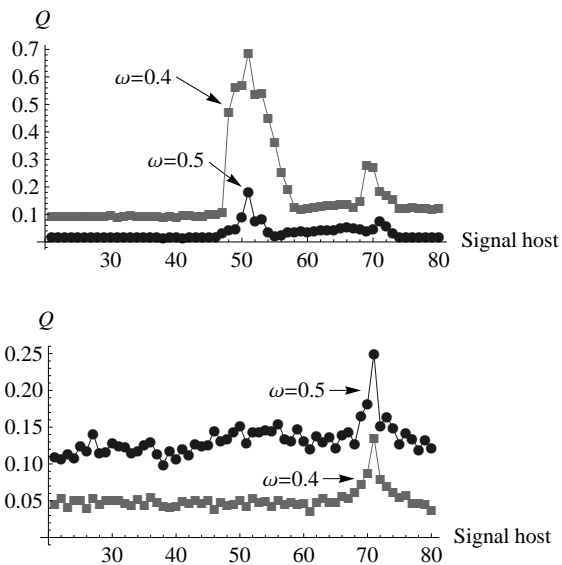}}
\vspace{-4cm} \caption{ When an impurity with relatively lower critical current (fast impurity) is introduced in an almost homogeneous array, Fourier coefficient dependence on the position of the external signal shows a resonance when the host node is near the impurity (upper plot). Total number of junctions is $99$, the critical currents $\alpha_{j}$ are chosen from a uniform distribution in the range $[0.98,1.02]$ except for the impurity $\alpha_{i}=0.8$ with $i=50$. Sample junction here is chosen $n=70$ but qualitatively, the result is independent of the size of the array and the number of sample junction if the transients has been elapsed. External dc current is $i_{j}=1.02$ for all $j$. Damping parameter and coupling constant are $\beta_{c}^{-1/2}=0.75$ and $k_{0}=0.25$, respectively. The external periodic input with amplitude $a_{m}=0.1$ with two different frequencies is imposed only on the $m$th junction and $m$ is varied from $20$ to $80$. Boundary conditions are absorbing i.e. the dc drive of boundary junctions is switched off to prevent the waves to re-enter into the array. In the lower plot results are shown for the array without impurity.} \vspace{-.0cm}
\label{fig1}
\end{figure}

Equation 2 describes the damped driven FK model\cite{fkbook}. We consider an almost homogeneous array into which has been introduced a single junction (the impurity) that has a critical current relatively lower than that of the other junctions. All the junctions then are driven by a constant current which is larger than their critical current. In the absence of the inductive coupling all the junctions would be in rotating state, and the impurity would rotate faster than the rest of the array.

We place the fast impurity at the site $i=50$ of a chain of $99$ junctions, and impose the periodic signal at site $m$. The critical currents of the other junctions are chosen from a uniform distribution in the range $[0.98,1.02]$. The response of the system to the pacemaker is probed via the Fourier coefficients $Q_{n}=\sqrt{R_{n}^2+W_{n}^2}$ according to\cite{perc}
\begin{eqnarray}
R_{n} =\frac{2}{T}\int_{0}^{T}\sin(\omega t) v_{n}(t) dt, \\
W_{n} =\frac{2}{T}\int_{0}^{T}\cos(\omega t) v_{n}(t) dt.
\end{eqnarray}
Here $v_{n}=d\theta_{n}/dt$ is the voltage of a sample junction (we take $n=70$ through the paper) and $T$ is the integration time. We then vary the location at which the periodic signal is imposed and record $Q_{n}$ as a measure of how the time course of the sample node is correlated with the external signal.

As is seen in Fig. 2, the system responds to the presence of a fast impurity by a sharp resonance when the nodes near the fast impurity host the pacemaker (There is also a boosted response when pacemaker is located on the sample junction, but this is trivial). The results show only minor variation in different trials; the results we quote are obtained by averaging over $50$ trials.  We also observe that changing the frequency affects the magnitude of the correlation maximum, but the behavior of the system is the same in that a resonance is seen when the signal is located on the impurity. Without an impurity the response of the system shows considerably more variability from trial to trail and upon averaging over the trials the response shows a maximum only when the signal is imposed on the sample node and its neighbors.

Study of the dependence of the response of the model on the frequency reveals why the resonances in Fig. 2 appear with different magnitudes. In Fig. 3 the Fourier coefficients $Q_{n}$ are plotted for a range of the frequencies of the periodic signal, with the other parameters the same as the Fig. 2. Again two cases are considered: when the signal is imposed on the fast impurity and when it is placed on another junction. The figure shows a resonant effect for certain values of the frequency (multiples of approximately 0.4 for both cases), but the resonance amplitude is considerably larger when the signal is located on the fast impurity. Note that since the critical currents of the impurity and the other junctions are different, in isolation they would have different resonant frequencies. But for the array (with linear couplings) whether the external signal is imposed on the impurity or on the other junctions, the resonant frequency would be identical and for all the values of frequencies the response of the array is larger if the signal is imposed on the impurity.

\begin{figure}[ht!]
\vspace{-0cm}
\centerline{\includegraphics[width=8.5cm]{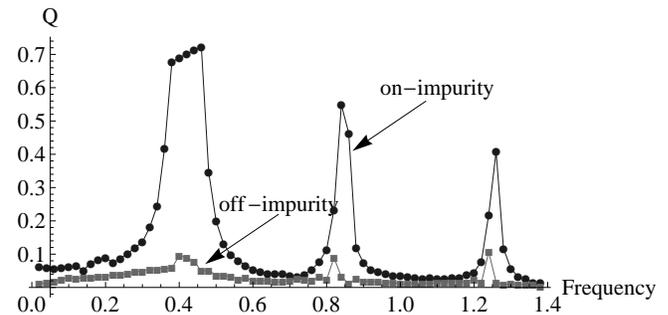}}
\vspace{-0cm} \caption{(a) Fourier coefficient vs. frequency of the external signal when the local periodic signal with amplitude $a_{m}=0.1$ is imposed on the impurity (dark gray circles) and on $60$th junction (light gray squares). Number of junctions is $99$ and the sample junction is chosen $n=70$ but again the result is independent of the position of sample junction and size of the array. Critical currents of all the junctions are chosen from a uniform distribution from the range $[0.98,1.02]$, except for the middle junction which has smaller critical current $\alpha_{i}=0.8$ with $i=50$. External dc current is $i_{j}=1.02$ for all $j$. Damping parameter and coupling constant are $\beta_{c}^{-1/2}=0.75$ and $k_{0}=0.25$, respectively. } \vspace{-0.0cm}
\label{fig4}
\end{figure}

The role of the fast impurity in creating a {\it location sensitive response} to the local inputs can be better understood by studying the I-V characteristic of the array. For a single junction it is known that the width of the Shapiro steps shows a Bessel function type relation to the amplitude of the signal\cite{bessel}, and for the small amplitude signals which we consider in this study, the width of the main Shapiro step has a linear relation with the amplitude of the periodic current. As noted before, in the array with inductive couplings, the I-V characteristic for all the junctions would be the same in the steady state regime (after transients have died out). We have shown the width of the main Shapiro step in the characteristic of the sampling junction  when the signal is imposed on the impurity and when on another junction in the array (Fig. 4). It can be seen the width of the main step for the junctions in the array shows a linear relation with signal amplitude when signal is imposed on the impurity. On the other hand, if the signal is located on another junction, no locking region is seen for small amplitude signals.

\begin{figure}[ht!]
\vspace{0cm}
\centerline{\includegraphics[width=9cm]{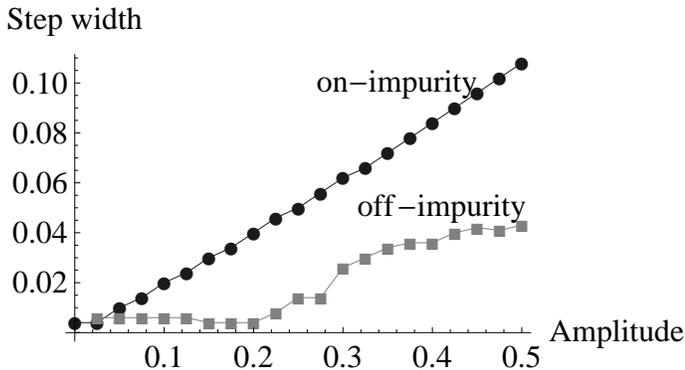}}
\vspace{0cm} \caption{ The width of main $1:1$ Shapiro step for an array when the local signal is imposed on the fast impurity (dark gray circles) and on the $60$th junction (light gray squares). Total number of junctions is $99$, sample junction here is chosen $n=70$ but the results are independent of the index of sample junction and the size of the array. Critical currents of all the junctions are chosen from a uniform distribution in the range $[0.98,1.02]$, except for the middle junction $\alpha_{i}=0.8$ with $i=50$. Damping parameter and coupling constant are $\beta_{c}^{-1/2}=0.75$ and $k_{0}=0.25$, respectively. } \vspace{-0.0cm}
\label{fig4}
\end{figure}

Figure 4 further shows that an off-impurity signal can entrain the array dynamics if its amplitude is larger than a critical value. For off-impurity signals with amplitude larger than a threshold, the critical current of the junction which receives the signal is decreased, so that the impurity loses its role as the fastest junction (the junction with largest value of voltage when isolated). In this case the entrainment of the array by a large amplitude off-impurity signal would be possible.

The behavior of the system described above, originates from the role of the fast component in generating solitary pulses consisting of a pair of kink-antikink (see Ref. \cite{impurity2}). When the periodic signal is imposed on the fast component, the rate of the nucleation of the kinks (which is proportional to the rate of the change of the phase of the junction) can be locked to the external frequency. The solitary pulses themselves move along the array and entrain the whole array after a transient time which grows with the size of the array. A signal that is imposed on the other junctions can entrain the nearby junctions but has no effect on the rate of solitary pulses which are being produced by the impurity; then a long range influence is not expected by off-impurity signals. We also note that the existence of solitary excitations of the FK model is crucial for existence of the behavior seen above: a ladder arrangement of Josephson junction is a counterexample in which the long range effect of local signals can not be seen in it. As we will show in the subsequent section, same behavior can be seen in other models while the model supports excitation which do not decay in space.

\section{Array of coupled FitzHugh-Nagumo oscillators}
While FitzHugh-Nagumo model is originally proposed as a simplification of the Hodgkin-Huxley equations\cite{FHN}, it has been widely used as a generic model for excitable systems and media and can be applied to a variety of systems\cite{FHN1}. Here we consider an array of diffusively coupled FHN-type oscillators:
\begin{eqnarray}
\nonumber
   \frac{dv_{i}}{dt}=v_{i}-v_{i}^3 /3-w_{i}+I_{i}-g (v_{i+1}+v_{i-1}-2v_{i}), \\
   \frac{dw_{i}}{dt}=0.08[2.5+2.5 \tanh (\eta v_{i})-w_{i}],
\end{eqnarray}
 $v_{i}$ and $w_{i}$ are the fast (voltage) and slow (recovery) variables, respectively. $I_{i}$ is the external current and $g$ is the coupling constant (synaptic strength). The model oscillators show type-I excitability and undergo an infinite period bifurcation in $I_{ext}=2/3$ when $\eta \gg 1$\cite{hva}. Inter-spike intervals (ISIs) are defined as the interval between two successive spikes when the neurons is spiking repeatedly, i.e. for $I_{ext}>2/3$, and decease with increasing external current.

 In an almost homogeneous array, we introduce again an impurity with higher rate of activity and impose a small amplitude periodic signal on one of the neurons in the array. We then record ISIs in a sample neuron. If the period of the external signal is large compared to the ISI of the neurons, the response of the system to the external signal can be probed by the extent of the variation of the ISIs in a sample neuron. Maximum and the minimum of the ISIs of a sample neuron in a period of the external signal is recorded in our experiment and the difference of them is reported as the amplitude of the response of the system. Results shown in Fig. 5 indicate that in presence of the impurity a very sharp resonance when the impurity (and its neighbors) host the signal. In the absence of the impurity the response of the array is trivial and shows a maximum when the signal is imposed on the sample neuron. As for the FK model, the result is independent of the size of the array and location of the sample neuron while the system is in the steady state.

\begin{figure}[ht!]
\vspace{-0cm}
\centerline{\includegraphics[width=9cm]{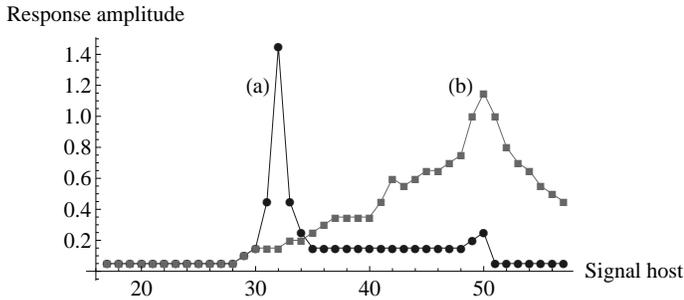}}
\vspace{-0cm} \caption{Amplitude of variation of inter-spike intervals is plotted vs. the position of the neuron on which a periodic signal is imposed. The chain consists of 63 neurons. In (a) a fast spiking impurity is in the middle (Dark gray circles). In (b) experiment is repeated in an array without impurity (light gray squares). External currents for all the neurons are chosen from a uniform distribution in the range $[1.18,1.22]$ except for the impurity which receives a larger input $I_{32}= 2.2$ to fire with a higher rate. Frequency and the amplitude of the external periodic signal are $a = 0.2$ and $\omega= 0.005$, respectively. The sample neuron is chosen $n=50$.} \vspace{-.0cm}
\label{fig4}
\end{figure}

\section{conclusion}

In conclusion, we have shown when a fast impurity is present in an array of inductively coupled Josephson junctions, locally imposed periodic signals show different levels of influence depending on whether they are imposed on the impurity or not. In such an array the fast impurity has a leading role as the source of solitary waves and even a small amplitude periodic signal when is imposed on the impurity can entrain the dynamics of the array. Otherwise, if the signal is imposed on the other junctions, the junctions of the array do not lock to the signal unless the amplitude of the signal is large.

The results are independent of the size of the array and for the large arrays, just transients are longer. Similar results are expected to be observed in the arrays which support non-decaying excitations. As an example which can find application in the studies on neuronal networks, we have considered an array of diffusively coupled FitzHugh-Nagumo oscillators. It has been shown that in such an array, a small amplitude signal can affect firing rate of all the neurons in array if it is imposed on the fast impurity.

\section*{Acknowledgement}
Author gratefully acknowledges J. P. Straley and M. Perc for reading the manuscript and giving useful comments.

\end{document}